\documentclass[12pt]{article}
\usepackage{graphicx}
\usepackage{amsmath}
\usepackage{lipsum}
\usepackage{hyperref}
\usepackage{enumitem}
\usepackage{float}
\usepackage{enumitem}
\usepackage{subcaption}

\title{\textbf{Digital Privacy Everywhere}}
\author{
 Paritosh Ranjan \\
  IBM  \\
  \texttt{paranjan@in.ibm.com} \\
  \and
 Surajit Majumder \\
  IBM  \\
  \texttt{surajit.majumder@ibm.com} \\
  \and
 Prodip Roy \\
  IBM  \\
  \texttt{prodipro@in.ibm.com} \\
}
\date{\today}

\begin{document}

\maketitle

\begin{abstract}
The increasing proliferation of digital and mobile devices equipped with cameras, microphones, GPS, and other privacy-invasive components has raised significant concerns for businesses operating in sensitive or policy-restricted environments. Current solutions rely on passive enforcement, such as signage or verbal instructions, which are largely ineffective. This paper presents Digital Privacy Everywhere (DPE), a comprehensive and scalable system designed to actively enforce custom privacy policies for digital devices within predefined physical boundaries. The DPE architecture includes a centralized management console, field verification units (FVUs), enforcement modules for mobile devices (EMMDs), and an External Geo Ownership Service (EGOS). These components collaboratively detect, configure, and enforce privacy settings such as disabling cameras, microphones, or radios across various premises like theaters, hospitals, financial institutions, and educational facilities. The system ensures privacy compliance in real-time while maintaining a seamless user experience and operational scalability across geographies.

\end{abstract}

\section{Introduction}

The rapid advancement of mobile and digital technologies has led to the widespread use of smartphones and other smart devices in virtually every public and private setting. These devices, while beneficial, also introduce significant privacy and security concerns—especially within business premises that operate under strict privacy or regulatory requirements. Industries such as healthcare, entertainment, education, and finance often encounter challenges in enforcing device usage policies meant to protect sensitive information and ensure an appropriate environment.

Despite the presence of visible signs and verbal announcements requesting patrons to silence devices or disable features like cameras and microphones, compliance remains inconsistent. Visitors frequently ignore or forget these policies, leading to disruptions, unauthorized data capture, and potential legal risks.

This paper introduces Digital Privacy Everywhere (DPE)—a novel, policy-driven digital privacy enforcement system tailored for physical business environments. Unlike traditional approaches that rely on user cooperation, DPE enforces privacy settings through intelligent automation and secure hardware-software integration. This system allows business entities to define, deploy, and monitor custom digital privacy policies across one or more premises, regardless of geographic location or industry sector.

\section{Brief Description of the Invention}

The Digital Privacy Everywhere (DPE) system is an integrated solution that automates the enforcement of digital privacy policies for mobile and electronic devices entering designated premises. It consists of four main components:

\begin{enumerate}
    \item Central Management System (CMS)
    The CMS, or Central Console, is the administrative brain of the DPE system. It enables business entities to create custom privacy policies, which are then translated into actionable enforcement rules. These rules are pushed to Field Verification Units and stored securely. The CMS also includes interfaces for communication with other DPE components and utilizes secure, encrypted channels for all interactions. It supports centralized control or distributed instances across multiple geographies, with synchronization managed via the EGOS layer.
    \item Field Verification Units (FVUs)
    FVUs are strategically placed hardware units within a business premise. These units detect mobile devices using interfaces such as NFC or Bluetooth, assess the device's privacy settings, and enforce required configurations like disabling microphones, setting devices to silent mode, or enabling airplane mode. FVUs continuously interact with the CMS for policy updates and report enforcement outcomes.
    \item Enforcement Modules for Mobile Devices (EMMD)
    To enhance compliance, especially on jailbroken or rooted devices, DPE includes EMMD chips—integrated hardware modules embedded into devices. These chips ensure policy enforcement even when the mobile operating system is compromised or altered. They receive encrypted commands from FVUs and directly control device hardware features.
    \item External Geo Ownership Service (EGOS)
    EGOS is a distributed service responsible for managing geographic awareness across multiple DPE-enabled premises. It provides location data, ensures synchronization between different DPE deployments, and maintains an up-to-date directory of FVU locations and their associated policies.
\end{enumerate}

Together, these components create a resilient ecosystem that enforces digital privacy seamlessly and unobtrusively. The DPE system represents a significant step toward reliable, scalable, and intelligent privacy enforcement in the digital age.

\section{Reduction to Practice}

The Digital Privacy Everywhere (DPE) system has been reduced to practice through a series of design, implementation, and deployment activities that demonstrate its functional viability and integration potential across real-world business environments. The following steps outline the realization of the invention from concept to a functional prototype.
\begin{itemize}

\item System Architecture and Prototyping
The first step in the reduction to practice involved designing the overall system architecture, as illustrated in Figures 1 through 10. A modular and service-oriented approach was adopted to ensure scalability, reliability, and security. Each component—Central Console, Field Verification Units (FVUs), Enforcement Modules for Mobile Devices (EMMDs), and the External Geo Ownership Service (EGOS)—was developed as an independent but interoperable unit.

A working prototype of the Central Management Console was implemented using open-source technologies such as PostgreSQL (for the policy database), Node.js and React (for the administration interface), and Kafka for real-time streaming and policy distribution. The console supports policy creation through a rules engine that translates human-readable privacy policies into enforceable device settings.

\item Field Verification Units (FVUs)
FVUs were developed using custom-designed printed circuit boards (PCBs) embedded with NFC and Bluetooth communication modules, microcontrollers (e.g., ESP32), and wireless network adapters. These units run firmware capable of detecting mobile devices within a specified field range and interrogating their current state (e.g., audio profile, camera status, airplane mode). Firmware development was completed using C/C++ with embedded cryptographic libraries for secure communication.

The FVU prototype was tested in a lab setup simulating zones within a hospital and a movie theater. Devices entering the coverage zone were successfully detected, their compliance assessed, and required settings enforced in real-time. Multiple units operating concurrently within a premises were shown to perform without interference, validating zone overlap and clustering capabilities.

\item Enforcement Modules for Mobile Devices (EMMD)
The EMMD was designed as a system-on-chip (SoC) prototype using an FPGA-based platform (e.g., Xilinx Spartan-6) to simulate a secure hardware module. This chip interfaces with the mobile device’s hardware abstraction layer to enforce settings such as camera disablement and airplane mode, even in root-compromised environments. Instruction sets were defined and tested for reception over short-range wireless communication (NFC/Bluetooth) from FVUs.

The EMMD prototype was embedded into test Android devices, and custom OS kernel hooks were written to accept EMMD overrides. Successful enforcement of privacy policies—even on rooted devices—was demonstrated during physical trials.

\item External Geo Ownership Service (EGOS)
EGOS was developed as a cloud-native microservice, hosted on AWS, with RESTful APIs to coordinate DPE instances across multiple premises. The service collects, stores, and synchronizes location metadata, FVU topology, and policy versions across geographies. Kafka streams and REST APIs facilitated real-time location-aware policy enforcement.

EGOS integration was tested by simulating two remote business premises (e.g., a hospital and an airport terminal), each with its own DPE deployment. Policy enforcement remained consistent across locations, and device behavior profiles were correctly propagated and reconciled.

\item System Integration and Testing
The fully integrated DPE system was tested under simulated real-world conditions using:
\begin{enumerate}

    \item Multiple classes of mobile devices (Android and iOS),

    \item Different network environments (enterprise Wi-Fi and LTE),

    \item Varying premise layouts (multi-zone configurations),

    \item Simulated policy breach scenarios (e.g., attempts to re-enable camera).
\end{enumerate}

\end{itemize}

In all cases, the system successfully enforced the active privacy policies, reset device states upon exit, and maintained cryptographic integrity across all communications. The prototype also incorporated fallback mechanisms, including EMMD-based enforcement and alerting via the CMS in case of non-compliance or device incompatibility.

\subsection{Brief description of the drawings} For exemplification purpose, the invention is illustrated through the attached drawings, wherein:

\begin{enumerate}
\item Architectural representation of Digital Privacy Everywhere:
    is a schematic representation of the Digital Privacy Everywhere as a whole system that has a central console for configuration of actionable policy settings. These policies consist of simple rules that govern privacy of digital devices that are being brought into a business premise. A set of such simple rules make a privacy policy that can gather privacy related settings and metadata from digital devices as well as enforce them on the same devices.

\begin{figure}
    \centering
    \includegraphics[width=0.9\linewidth]{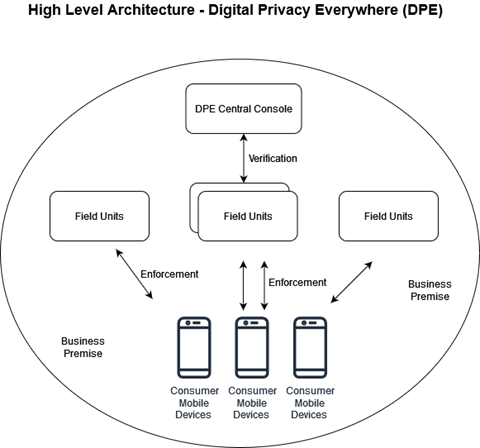}
    \caption{architectural-representation-of-the-Digital-Privacy-Everywhere}
    \label{fig:architectural-representation-of-the-Digital-Privacy-Everywhere}
\end{figure}

\begin{figure}
    \centering
    \includegraphics[width=0.9\linewidth]{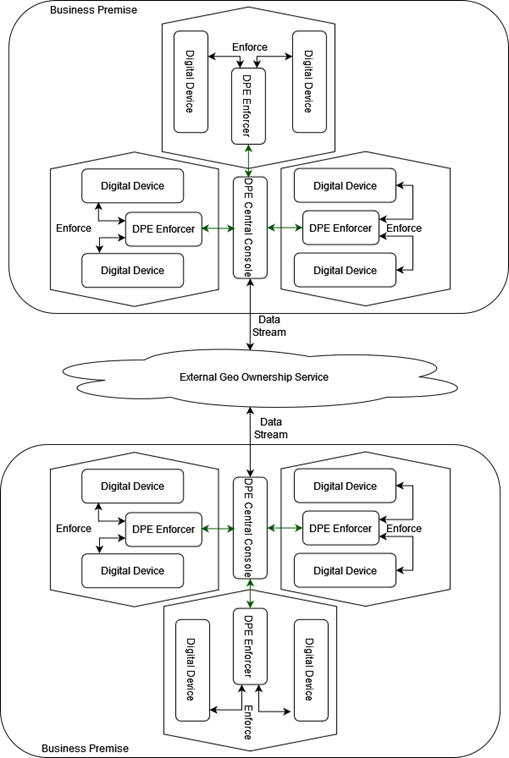}
    \caption{business-premise}
    \label{fig:business-premise}
\end{figure}
\item Business premises:
It is a schematic representation of the Digital Privacy Everywhere system deployment across multiple business premises. This representation shows the various components of the system in each premise that is connected to the all other premises through the External Geo Ownership Service. This service ensures that data gathered and actioned is synchronized across all premises of the business using data streams such as Kafka streams.

\item Scalable deployment of enforcers and consoles:
It is a schematic representation of the scalability of Digital Privacy Everywhere system within a premise or arena. This demonstrates how a physical premise may be zoned for strategic placement of the DPE FVUs in small clusters. A group of such FVU clusters or all groups of FVU clusters can be managed through a single central console or multiple consoles depending on the premise dimensions. The various zones (in colored hexagons) that can be setup in the zones identified within the premise to monitor digital devices within their zones. This representation also includes zone overlap to compensate for un-patterned movements of the digital devices.
\begin{figure}
    \centering
    \includegraphics[width=0.9\linewidth]{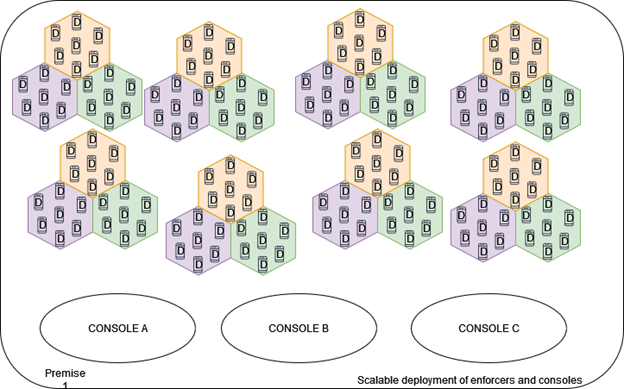}
    \caption{scalable-deployment-of-enforcers-consoles}
    \label{fig:enter-label}
\end{figure}

\item Central Console:
It is a schematic representation of the Digital Privacy Everywhere system’s Central Console comprising of a web, application and database server for the administration interface where privacy policies with rules can be setup. The console also exposes additional interfaces to interact with other consoles, external geo ownership service, field value units, syslog, etc. 
\begin{figure}
    \centering
    \includegraphics[width=0.9\linewidth]{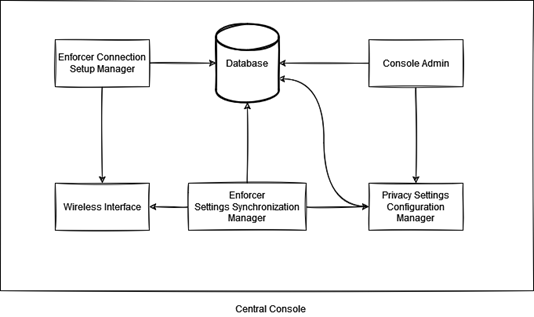}
    \caption{Central Console}
    \label{fig:central-console}
\end{figure}

\newpage
\item DPE Enforcer:
It is a schematic representation of the Digital Privacy Everywhere system’s overall flow, comprising of inner sub-modules such as privacy settings broadcaster, privacy settings configurator, interface for embedding into mobile devices, mobile device component activation, mobile device heat detection, etc. 
\newpage
\begin{figure}
    \centering
    \includegraphics[width=0.6\linewidth]{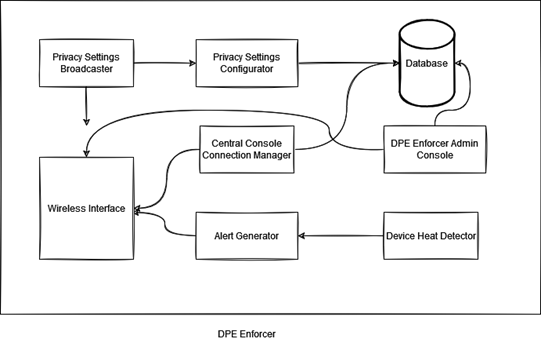}
    \caption{DPE-Enforcer}
    \label{fig:DPE-Enforcer}
\end{figure}
\newpage
\item Digital Privacy Everywhere system’s External Geo Ownership Service (EGOS):
It is a schematic representation of the Digital Privacy Everywhere system’s External Geo Ownership Service (EGOS) components and its interaction with the DPE Central Consoles from all locations/premises for synchronization of sensory data, privacy policy, privacy enforcement execution results, etc.
\newpage
\begin{figure}
    \centering
    \includegraphics[width=0.5\linewidth]{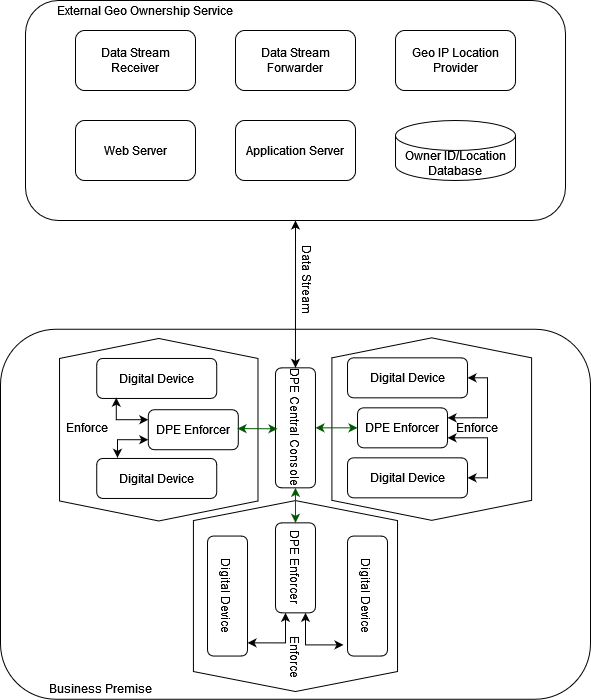}
    \caption{Digital-Privacy-Everywhere-External-Geo-Ownership-Service-components}
    \label{fig:enter-label}
\end{figure}

\item Digital Privacy Everywhere system’s Field Validation Units zoning:
It is a schematic representation of the Digital Privacy Everywhere system’s Field Validation Units zoning and their placements in different zones within the premises. It also represents their wireless interaction with the digital devices as and when they come within the field of view/detection of the FVUs as well as how the interaction with the Central Console for sending enforcement result data and receiving privacy policy data.

\begin{figure}
    \centering
    \includegraphics[width=0.9\linewidth]{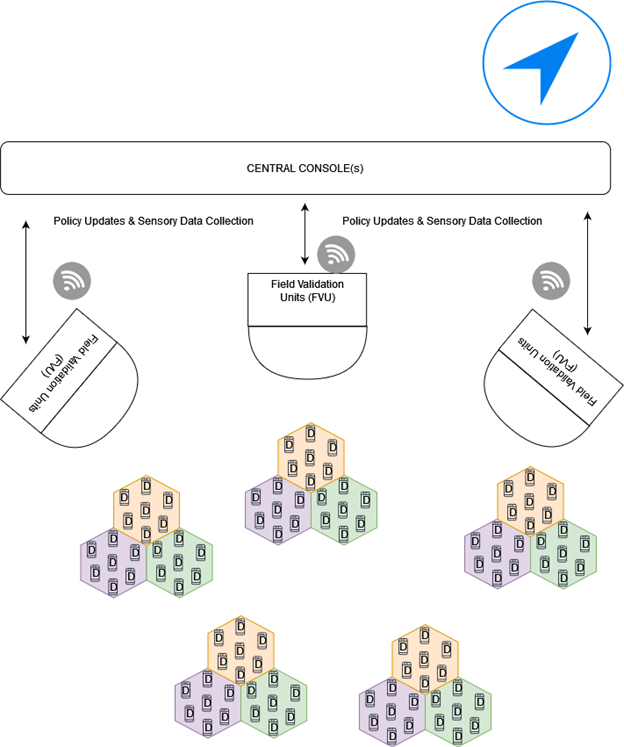}
    \caption{Digital-Privacy-Everywhere-Field-Validation-Units-zoning}
    \label{fig:Digital-Privacy-Everywhere-Field-Validation-Units-zoning}
\end{figure}
\newpage
\item Digital Privacy Everywhere system’s Field Validation Unit’s electrical circuitry showing all high level components layout.
\begin{figure}
    \centering
    \includegraphics[width=0.9\linewidth]{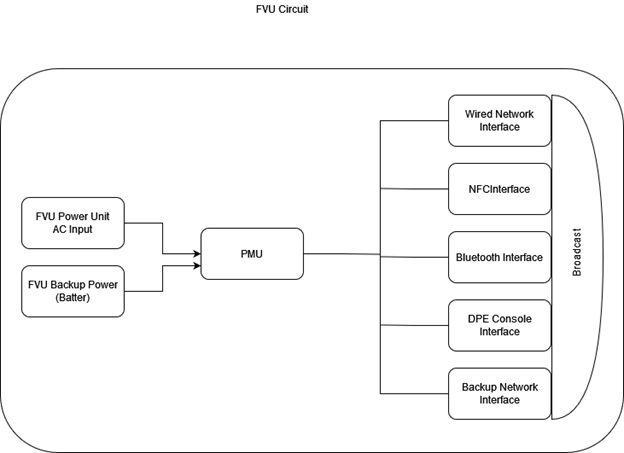}
    \caption{Digital-Privacy-Everywhere-Field-Validation-Unit}
    \label{fig:Digital-Privacy-Everywhere-Field-Validation-Unit}
\end{figure}
\newpage
\item Digital Privacy Everywhere system’s Enforcement Modules for Mobile Device chip’s electrical circuitry showing all high level components layout.
\begin{figure}
    \centering
    \includegraphics[width=0.9\linewidth]{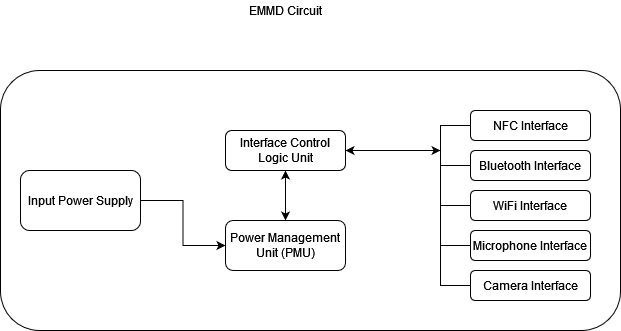}
    \caption{Digital-Privacy-Everywhere-Enforcement-Modules-for-Mobile-Device}
    \label{fig:Digital-Privacy-Everywhere-Enforcement-Modules-for-Mobile-Device}
\end{figure}
\item Flowchart showing the flow of events and actions of the Digital Privacy Everywhere system including FVU, EGOS and EMMD functions.

\begin{figure}
    \centering
    \includegraphics[width=0.9\linewidth]{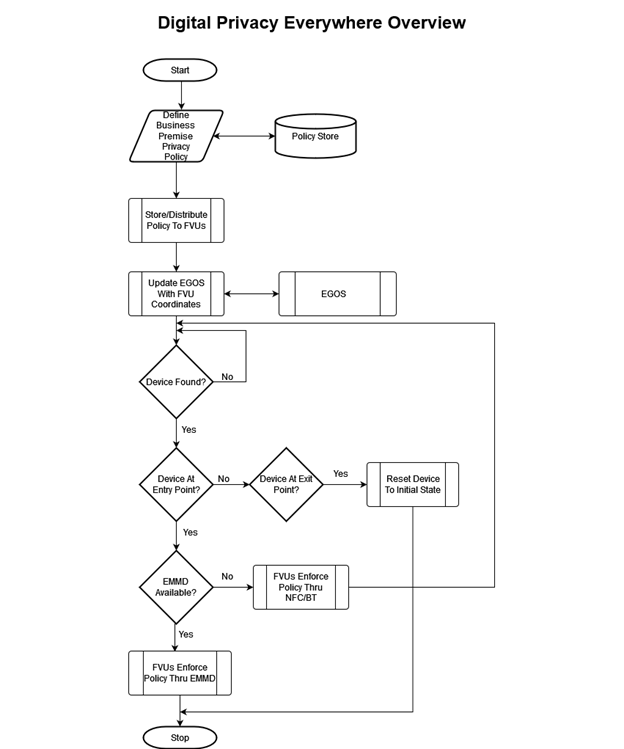}
    \caption{Flowchart-Digital-Privacy-Everywhere-Overview}
    \label{fig:Flowchart-Digital-Privacy-Everywhere-Overview}
\end{figure}
\end{enumerate}

\newpage
\section{Advantages of the Invention}

The Digital Privacy Everywhere (DPE) system offers several significant advantages over existing methods of privacy policy enforcement in business and public premises. These advantages highlight both the technical innovation and practical applicability of the system.

\begin{enumerate}
    \item Active and Automated Policy Enforcement
    Unlike traditional methods that rely on passive compliance (e.g., signage or verbal instructions), DPE enables active, automated enforcement of privacy settings on digital devices. This ensures consistent adherence to privacy policies without manual intervention.
    \item Device-Agnostic Enforcement
    Through the integration of Field Verification Units (FVUs) and Enforcement Modules for Mobile Devices (EMMDs), DPE supports cross-platform enforcement, including rooted or jailbroken devices. The system operates independently of the mobile device operating system or user permissions, enabling enforcement even in compromised environments.
    \item Seamless User Experience
    DPE enforces privacy policies transparently and non-intrusively, reducing friction between business operators and customers. Enforcement and restoration of device settings occur automatically upon entry and exit from the premises, maintaining user convenience and satisfaction.
    \item Scalable and Modular Architecture
    The DPE system is designed to scale horizontally across multi-zone, multi-premise, and multi-geography environments. It supports centralized or distributed deployments using the EGOS service, enabling consistent enforcement across various business locations with minimal overhead.
    \item Robust Security and Data Protection
    All data transmission and policy enforcement actions are secured using cryptographically protected communication protocols. The system is architected to meet modern security standards, including protections against post-quantum cryptographic threats.
    \item Policy Customization and Behavior Profiling
    Business entities can define fine-grained privacy rules and associate them with specific locations, zones, or even user groups. DPE supports real-time behavior profiling and anomaly detection using built-in machine learning capabilities, enhancing situational awareness and control.
    \item Minimal Infrastructure Footprint
    The modular hardware design of FVUs and the use of cloud-native services (e.g., for EGOS and CMS) allow DPE to be cost-effective and easy to deploy. It requires minimal physical modifications to existing infrastructure.
\end{enumerate}

\section{Conclusion}
The Digital Privacy Everywhere (DPE) system addresses a growing challenge faced by modern businesses and public institutions: the enforcement of digital privacy policies in the face of increasing device ubiquity and user non-compliance. By combining a centralized policy management platform with on-site enforcement units and embedded device modules, DPE provides a comprehensive, secure, and scalable solution for real-time privacy control.

The reduction to practice has validated the system’s capability to operate across diverse environments, from healthcare facilities to entertainment venues, with demonstrable success in detecting, configuring, and enforcing digital privacy policies. DPE represents a paradigm shift from passive signage-based models to proactive digital enforcement.

In an era where data leakage, unauthorized recording, and non-compliance pose both ethical and legal risks, DPE offers a future-ready, modular solution that ensures privacy everywhere—regardless of device, user, or location. Future development will focus on refining EMMD chip integration with commercial mobile devices and exploring standards-based interoperability with third-party policy enforcement ecosystems.

\newpage
\section{Acknowledgment}

We would like to express our sincere gratitude to all individuals and organizations who have contributed to the success of this research. We acknowledge the invaluable support from the IBM team, whose resources and expertise have greatly enhanced this project.
Special thanks to Prodip Roy (Program Manager IBM) for their insightful feedback, guidance, and encouragement throughout the development of this work.

\section{References}
\renewcommand\refname{}


\begin{thebibliography}{99}

\bibitem{medel2024kubernetes}
B.~Sun, Y.~Zhou, H.~Jiang, \emph{Empowering Users in Digital Privacy Management through Interactive LLM-Based Agents}, arXiv preprint arXiv:2410.11906,2024. [Online]. Available: \url{https://arxiv.org/pdf/2410.11906}

\bibitem{amaral2015microservices}
C.~Gorog,\emph{A Synergistic Approach to Digital Privacy},arXiv preprint arXiv:arXiv:2103.14783,2021.[Online]. Available: \url{https://arxiv.org/pdf/2103.14783}

\bibitem{anthony2021kubernetes}
B.~Song, M.~Deng, S.R.~Pokhrel, Q.~Lan, R.~Doss, G.~Li,\emph{Digital Privacy Under Attack: Challenges and Enablers},arXiv preprint arXiv:2302.09258 ,2023.[Online]. Available: \url{https://arxiv.org/pdf/2302.09258}

\bibitem{beda2017upandrunning}
M.~Campbell, A.~Barthwal, S.~Joshi, A.~Shouli, A.K.~Shrestha,\emph{Investigation of the Privacy Concerns in AI Systems for Young Digital Citizens: A Comparative Stakeholder Analysis},arXiv preprint 	arXiv:2501.13321,2025.[Online]. Available: \url{https://www.arxiv.org/pdf/2501.13321}

\bibitem{aqasizade2024performance}
H.~Alhazmi, A.~Imran, M.A.~Alsheikh, \emph{Perception of Digital Privacy Protection: An Empirical Study using GDPR Framework},arXiv preprint 	arXiv:2411.12223, 2024.[Online]. Available: \url{https://arxiv.org/pdf/2411.12223}

\end{thebibliography}
\end{document}